\documentstyle[multicol,epsfig,prl,aps]{revtex}
\tighten
\draft
\begin{document}
\title{Normal Helium 3: a Mott-Stoner liquid.} 
\author{Antoine Georges} 
\address{ Laboratoire de Physique Th\'{e}orique de l'Ecole 
Normale Sup\'{e}rieure,\cite{auth1}
24 rue Lhomond, 75231 Paris Cedex 05, France}
\author{Laurent Laloux}
\address{Laboratoire de Physique Th\'{e}orique et Hautes 
Energies,\cite{auth2}
4 Place Jussieu, 75252 Paris Cedex 05, France}
\date{October 9, 1996}
\maketitle
\begin{abstract}
A physical picture of normal liquid $^3$He, which accounts for
both ``almost localized'' and ``almost ferromagnetic'' aspects,
is proposed and confronted to experiments.
\end{abstract}
\pacs{ PACS numbers: 67.55.-s, 75.10.Lp, 71.30.+h }

\begin{multicols}{2}
\narrowtext
Helium 3 is a liquid of strongly interacting fermionic atoms.
As pressure is increased between $p=0$ and $34$bars,
the compressibility
is drastically reduced from $\kappa/\kappa_0 \simeq .27$ to
$\kappa/\kappa_0 \simeq .066$, while the effective mass 
(specific heat coefficient) and magnetic susceptibility are enhanced 
from $m^*/m\simeq 2.8$, $ \chi/\chi_0\simeq 9.2$
to $m^*/m\simeq 5.8$, $\chi/\chi_0\simeq 24$ ($m$ is the mass of free
$^3\mbox{He}$ atoms, and $\kappa_0, \chi_0$ refer to the free Fermi gas
at the same pressure) \cite{CL}.
This is clearly the signal of the increasingly important effect of the
interatomic interaction, which has a strong repulsive hard-core.
Given the difficulties of a full quantitative
treatment of such a problem, more phenomenological
theoretical descriptions have been sought.

There are mainly two different, and seemingly contradictory,
physical pictures that have emerged over the years. 
In the ``almost ferromagnetic'' approach \cite{PARA,PN}, 
the liquid is viewed as being increasingly close to a 
ferromagnetic instability as pressure is increased and the 
large susceptibility is explained as a Stoner enhancement 
$\chi=\chi_0/(1-I\chi_0)$. Being very close to an instability,
critical spin-fluctuation modes (``paramagnons'' (PM)) 
must be taken into account beyond the mean-field Stoner description. 
These modes have been claimed to be essential to explain, for example,
the low-temperature dependence of the susceptibility \cite{BMF}. 
They also provide a logarithmic increase of the 
effective mass (though in rather mediocre quantitative agreement with 
experimental values). The ``almost ferromagnetic'' PM picture 
has some severe limitations however, particularly in failing to explain 
the strong reduction of compressibility \cite{PN}.

This reduction is one of the main motivation for viewing instead the 
liquid as ``almost localized'', {\it i.e} becoming more and more 
``solid-like'' with pressure, as first proposed by Anderson 
and Brinkman \cite{AB}, and extensively 
developed by Vollhardt \cite{DV}. 
The essential physics behind that picture is that of localization 
by repulsive interactions, in the sense of Mott.  
As explained below however, the simplest 
implementation of the quasi-localized picture leads to an incorrect 
description of the magnetic correlations and spin-fluctuation properties.  

In this letter, we would like to propose a novel description of liquid 
$^3\mbox{He}$, which retains the proximity to Mott localization 
as a central notion, but reintroduces a more accurate description of 
the spin-spin  correlations, of dominantly {\it ferromagnetic} nature.
In our  picture, in addition to being ``almost localized'', 
the liquid is also close to a ferromagnetic instability
(see also \cite{zazie}), 
but not in a critical regime (contrary to PM theory).
For this reason, liquid $^3\mbox{He}$ is viewed in our picture 
as a ``Mott-Stoner'' liquid.  
          
The quasi-localized picture was first implemented quantitatively 
\cite{DV,PN} 
by considering a lattice-gas model and modeling  
the hard-core repulsion as a Hubbard interaction:
\begin{equation}
H_1 = - \sum_{ij,\sigma} t_{ij}
c^{+}_{i\sigma} c_{j\sigma} + U\sum_{i} n_{i\uparrow} n_{i\downarrow}
\label{hubbard}
\end{equation}
This model was then treated using the Gutzwiller approximation (GA), 
which yields a  Mott transition at half-filling 
($n=1$) between a Fermi-liquid phase 
for $U<U_c$ and a localized Mott insulating phase for $U>U_c$.
Close to the transition, the compressibility vanishes as $U_c-U$
and the effective mass diverges as $1/(U_c-U)$, in qualitative
similarity with the behavior of liquid $^3\mbox{He}$ for increasing pressure. 
In the simplest description, the lattice-gas is constrained to 
be at half-filling, and the GA expression for $m^*/m$ can be fitted to the
experimental result in order to extract 
the single parameter $u(p)=U(p)/U_c$, with $u(p)$ an increasing 
function of $p$.  (More refined formulations \cite{SGRUV} also allow 
a variable filling factor,
and consider a trajectory $u=u(p)$, $n=n(p)$ in the $(u,n)$ plane).
Given $u(p)$, the calculated GA compressibility
is found to be in reasonable agreement with experiment.
It is interesting in this respect to consider the dimensionless ratio
$R_{\kappa}\equiv (\kappa/\kappa_0)(m^*/m)$, which is predicted to
reach a finite value at the transition $R_{\kappa}^{GA}(U_c)\simeq .25$.
Experimentally, this ratio does saturate at high pressure at a value
$R_{\kappa}^{exp}\simeq .38$. 

Turning to magnetic properties, the GA also leads to a divergent 
susceptibility $\chi\sim 1/(U_c-U)$, and thus to a finite ``Wilson ratio'' 
$R_W\equiv 1/(1+F_0^a)=(\chi/\chi_0)/(m^*/m)$. 
Experimentally, this ratio 
has a very weak dependence on pressure, 
with $R_W^{exp} (p=34\mbox{bars}) \simeq 4.1$, close to $R_W^{GA} (U_c)$.
 This agreement was originally viewed \cite{DV} as one of
the main success of the 
approach and interpreted as evidence that the susceptibility enhancement
could be entirely due to the incipient localization \cite{AB}, 
responsible for the effective mass enhancement.
However, we would like to point out that this divergence of 
the uniform susceptibility is {\it an artefact of the GA}, 
rather than a genuine feature of the Hubbard model. Indeed, in this model, 
the superexchange mechanism produces a nearest-neighbor 
{\it antiferromagnetic} exchange (of order $J\simeq \epsilon_F^2/2U$
\cite{EPS} for large 
enough $U$). On physical grounds, one expects this magnetic exchange to 
cutoff the divergence of the uniform susceptibility, which should remain 
{\it finite}, of order $\chi\simeq 1/J$ through the transition and 
in the localized phase. More accurate treatments of the Mott transition 
within a dynamical mean-field of the Hubbard model  
based on the limit of large lattice coordination 
fully support this view \cite{LISA}. Given the Fermi 
energy of liquid $^3\mbox{He}$, 
the magnetic superexchange would induce 
antiferromagnetic correlations on the scale of $J\simeq 350$mK, which 
is physically unrealistic and yields a too small
susceptibility enhancement $\chi/\chi_0\simeq \epsilon_F/J\leq 7$.
This has actually an even more drastic consequence, namely that 
the ground-state of the half-filled Hubbard model is in fact   
an antiferromagnetic insulating solid, rather than a paramagnetic liquid 
(except if a very large lattice frustration is 
introduced \cite{LISA}).
In the GA, the superexchange is neglected altogether so that these 
difficulties are simply overlooked.
The overestimate of short-range antiferromagnetic correlations 
is actually not 
entirely due to the lattice description of the system. Variational 
treatments in continuum space using Jastrow-Slater wave functions 
and a realistic interatomic potential, also suffer 
from similar problems (when confronted, e.g to neutron data)\cite{CL,BL}. 
This is because, as in the Hubbard model, the emphasis is put mainly on 
{\it only one} of the effects of the hard-core, namely the avoidance 
of double occupancy.    

Here, we suggest that the original formulation of 
the quasi-localized picture must be modified 
in order to account for the correct scale and type of magnetic 
correlations in the liquid. A very simple way 
to achieve this, while remaining in the framework of a lattice-gas 
model as above, is to introduce explicitly an additional nearest-neighbor 
{\it ferromagnetic} exchange,leading to the two-parameter model:
\begin{equation}
H_2 = - \sum_{ij,\sigma} t_{ij}
c^{+}_{i\sigma} c_{j\sigma} + 
U\sum_{i} n_{i\uparrow} n_{i\downarrow} - 
{{I}\over{z}} \sum_{<ij>} \vec{S}_i\cdot\vec{S}_j
\label{ham}
\end{equation}
The overall magnetic exchange $\tilde{I}\equiv I-J \simeq I-\epsilon_F^2/2U$, 
can have a priori an arbitrary sign,
and will be determined below from experimental considerations.
We shall use the dynamical mean-field approximation, formally exact in 
the limit of large lattice coordination $z\rightarrow \infty$,
to analyze the behavior of this model
(for an extensive review and technical details see Ref. \cite{LISA}).
In this limit, the inter-site magnetic interaction can be
treated at the (static) mean-field 
level (in contrast, the local 
Hubbard interaction requires a full dynamical treatment). 
Hence, the dynamical magnetic susceptibility 
$\chi(\vec{q},\omega)$ is readily obtained from that of the Hubbard model 
$\chi_H(\vec{q},\omega)$ at the same value of $U$ as:
\begin{equation}
\chi(\vec{q},\omega)^{-1}\,=\,\chi_H(\vec{q},\omega)^{-1} 
- I \Delta (\vec{q})
\label{chigen}
\end{equation}
where $\Delta(\vec{q})$ is the Fourier transform of the 
nearest-neighbor 
connectivity matrix: $\Delta(\vec{q})\equiv {{1}\over{z}}\sum_{j=1}^{z} 
\exp{(i\vec{q}\cdot\vec{R_j})}$. 
\begin{figure}[hbt]
 \parbox{\textwidth}{\epsfig{file=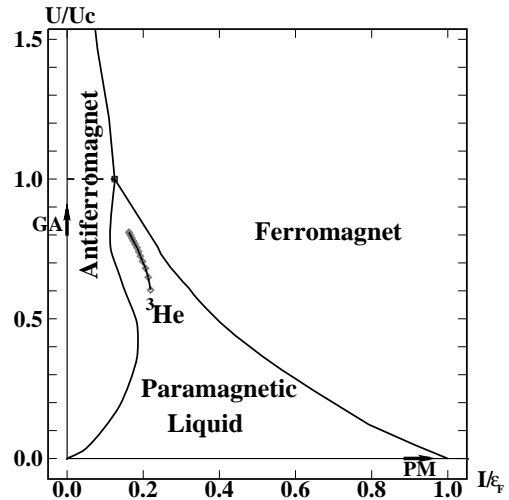,width=7cm,angle=270}} \\
\caption[0]{Phase diagram of model (\ref{ham}). The squares indicate
the fit to  liquid $^3$He data as pressure is increased from 
$p=0,3,6,\cdots, 33$bars. Also indicated by arrows are the corresponding
trajectories in the ``almost localized'' (GA) and the ``almost
ferromagnetic'' (PM) pictures.}
\label{FIG1}
\end{figure}
In Fig.1, we display the zero-temperature phase diagram of this model at 
half-filling, in the $(U,I)$ plane \cite{DOS,PHASE}.
We shall mainly focus in the following on the Fermi liquid phase 
which is found when the Hubbard repulsion is smaller than a critical 
value associated with Mott localization ($U<U_c\simeq 3.7 \epsilon_F$), 
and the ferromagnetic exchange $I$ is in the intermediate 
range $I_c^{AF}(U)<I<I_c^{F}(U)$. We emphasize that this phase does not 
display any kind of magnetic long-range order (the overall magnetic 
exchange $\tilde{I}$ being in an intermediate coupling regime, it is always 
successfully opposed by kinetic energy effects). Stabilizing such a 
phase is one of the primary motivation of our approach. 
(Note that, as expected, the half-filled pure Hubbard model
($I=0$) is ordered antiferromagnetically for all values of $U$ at $T=0$ in 
our treatment).
When $I$ is too small ($I<I_c^{AF}(U)$), the antiferromagnetic 
superexchange induced by the Hubbard repulsion takes over, and 
the liquid orders  antiferromagnetically, while for large 
$I>I_c^F(U)$, it orders ferromagnetically.
The Mott localized state for $U>U_c$ is always
magnetically ordered at $T=0$, either antiferromagnetically for 
$I<I_c^{AF}$ or ferromagnetically for $I>I_c^F$,
with $I_c^{AF}(U)=I_c^F(U)=J\simeq \epsilon_F^2/2U$ for $U>U_c$.

Simple estimates of the critical couplings $I_c^F, I_c^{AF}$ can be 
obtained for small $U$, and $U$ close to $U_c$\cite{EPS}. 
To first order in $U$, the Hubbard model static susceptibility reads: 
$\chi_H(\vec{q}=0)^{-1}=\chi_0^{-1}-U\equiv\epsilon_F-U$, 
and $\chi_H(\vec{Q})^{-1}=U$ for the antiferromagnetic 
wavevector $\vec{Q}=(\pi,\cdots,\pi)$. 
Inserting those into Eq.(\ref{chigen}), one obtains for small $U$: 
$I_c^F=\epsilon_F-U+O(U^2)$ and $I_c^{AF}=U+O(U^2)$. 
Close to the Mott transition, 
the static susceptibility of the Hubbard model in the dynamical mean-field 
approach is well approximated by the form \cite{LISA}: 
$\chi_H(\vec{q},\omega=0)^{-1}\simeq \lambda(\vec{q})\epsilon_F^* + 
J\Delta(\vec{q})$.
In this expression, $\lambda(\vec{q})$ has a rather weak 
$\vec{q}$-dependence (with $\lambda(\vec{0})= 1$), 
while $\epsilon_F^*$ is the low-energy effective 
Fermi scale $\epsilon_F^*\simeq .66 Z\epsilon_F$, 
with $Z$ the quasi-particle residue which
vanishes as the Mott transition is reached at half-filling: 
$Z \simeq  .9(1-U/U_c)$. Hence, the static 
$\vec{q}$-dependent susceptibility of (\ref{ham})
is reasonably approximated for $U \simeq U_c$ by:
\begin{equation}
\chi(\vec{q},\omega=0)^{-1}\,\simeq\, \lambda(\vec{q})\epsilon_F^* + 
(J-I)\,\Delta(\vec{q})
\label{chistat}
\end{equation}      
Using this expression, one obtains the estimates for $U$ close to $U_c$: 
$I_c^F\simeq J+\epsilon_F^*\simeq J+.16(U_c-U)$, 
$I_c^{AF}\simeq J-\lambda(\vec{Q})\epsilon_F^*$. 

We now focus on the behavior of various physical quantities in the 
strongly correlated liquid phase, and on the application to the physics 
of liquid $^3\mbox{He}$. In this phase, the single-particle self-energy, 
and the density-density 
response function are {\it unaffected} by the coupling $I$ within the 
dynamical mean-field treatment, so that the effective mass and 
compressibility are functions of $U$ only and coincide 
with those of the pure Hubbard model. Close to the Mott transition, 
the effective mass diverges as $m^*/m=1/Z\simeq 1.1/(1-U/U_c)$, while the 
compressibility vanishes as $\kappa/\kappa_0\simeq .66(1-U/U_c)$, 
which is qualitatively 
similar to the GA (in contrast with 
$\chi\simeq 1/(\epsilon_F^*+J-I)$).        
Following the original phenomenological spirit of the quasi-localized 
approach \cite{DV}, one can determine the pressure dependence of the 
two effective parameters $U(p)$, $I(p)$ by fitting the experimental 
results for two physical quantities. 
We have chosen to use the effective 
mass (specific heat) data to determine $U(p)$, and then
to extract $I(p)$ from the susceptibility. For each pressure, liquid 
$^3\mbox{He}$ thus corresponds to a point indicated on the phase diagram 
of Fig.1. 
The resulting trajectory is seen to approach {\it both} Mott localization 
$U=U_c$ and the ferromagnetic phase boundary $I=I_c^F(U)$ .
While the effective mass enhancement is entirely associated with 
the on-site repulsion $U$, the proximity of the ferromagnetic phase 
boundary is crucial to account for the observed magnitude of the 
susceptibility enhancement: at the value of $U$ corresponding to the highest 
pressures, the susceptibility enhancement of the pure Hubbard model would 
be $\chi_H/\chi_0\simeq 5.4$, about 4.5 times too small. 
In the present approach, the saturation of the Wilson ratio 
$R_W=1/(1+F_0^a)$ precisely 
reflects the fact that both instabilities are approached. Indeed, using 
Eq. (\ref{chistat}) we have, close to $U_c$: 
$\chi/\chi_0 = \epsilon_F/(\epsilon_F^*+J-I) = 
\epsilon_F/(I_c^F-I)$, 
so that $R_W=(\chi/\chi_0)/(m^*/m)\simeq \epsilon_F^*/(I_c^F-I)$, 
and hence $U_c-U(p) \simeq 4R_W (I_c^F-I(p))$, 
indicating a linear trajectory towards the multicritical point at the 
highest pressure. The total magnetic exchange 
$\tilde{I}\equiv I-J\simeq (.66R_W-1)/\chi
\simeq 1.6/\chi$ 
obtained from our approach is {\it ferromagnetic} ($\tilde{I}>0$) 
and of the order of $300$mK.
This is precisely the typical energy gained by including short-range 
magnetic correlations in variational calculations \cite{CL,BL}. 
The ferromagnetic sign is consistent with an instability of the liquid 
towards a {\it triplet} superfluid phase at low temperature. 
Having determined $U(p), I(p)$, we have compared 
the compressibility computed for our model to experiment.
Excellent agreement is found at low pressure, while calculated
values at high pressure are too large by approximately a factor of $2$.
The ratio $R_{\kappa}\equiv(\kappa/\kappa_0) (m^*/m)$ is predicted to
saturate at high pressure as observed experimentally
(with $R_{\kappa}(U_c)\simeq .73$, 
while $R_{\kappa}^{exp} (p=34\mbox{bars})\simeq .38$). 

We would now like to compare and contrast the physical picture 
proposed here to that of the ``almost ferromagnetic'' PM 
description \cite{PARA,PN}. We first evaluate the dimensionless parameter 
$r\equiv (\tilde{I}_c^F-\tilde{I})/\tilde{I}_c^F$  measuring 
the distance to the ferromagnetic critical boundary. From above, 
we find $r\simeq 1.5/R_W$, 
which varies from $r\simeq.46$ at low 
pressure to $r\simeq.36$ at high pressure and is thus never very small 
in the present approach (in contrast $r\simeq .11$ to $.042$ in PM 
theory). Hence, the ferromagnetic exchange may be treated within Stoner 
mean-field theory, with no significant effect of the long-wavelength 
PM fluctuations. This justifies {\it a posteriori} our 
treatment of this coupling within the large-connectivity limit. 
At low-energy, we have a liquid of quasi-particles 
characterized by the effective Fermi scale $\epsilon_F^*$. 
In the absence of any magnetic exchange ($\tilde{I}=0$), 
the susceptibility of this gas would be of 
order $\chi_{qp}\simeq 1/\epsilon_F^*$. The actual susceptibility 
$\chi=1/(\epsilon_F^*-\tilde{I})$ is correctly given by Stoner expression, 
with an effective Stoner enhancement $S_{eff}=\chi/\chi_{qp}\simeq .66R_W$.
$S_{eff}$ depends weakly on pressure, and
measures the fraction of the total susceptibility enhancement 
due to the exchange (in contrast, 
$S\equiv \chi/\chi_0 = 1.5 S_{eff}(m^*/m)$ 
is a combination of exchange and localization effects and strongly
depends on pressure).
These remarks 
also imply that there is no significant enhancement of the effective 
mass due to ferromagnetic spin fluctuations (in contrast with the 
logarithmic effect of PM theory), which leaves the estimate 
$m^*/m \simeq 1.1/(1-U/U_c)$ used above essentially unaffected. From 
Eq.(\ref{chistat}) and the fact that $\tilde{I}>0$, one sees that 
the susceptibility is peaked around $\vec{q}=0$. 
For low $|\vec{q}|\equiv q\ll k_F$ and 
$\omega\ll qv_F^*$ (with 
$v_F^*=Z v_F$ the effective Fermi velocity), we can approximate the 
dynamical susceptibility by (neglecting all other residual interactions 
between quasiparticles apart from the exchange):
\begin{equation}
\chi(\vec{q},\omega)^{-1}\simeq \epsilon_F^* \left( 
1-{{\tilde{I}}\over{\epsilon_F^*}} + b {{q^2}\over{k_F^2}} 
-i a {{\omega}\over{q v_F^*}} \right)
\label{chidyn}
\end{equation}
  From this expression, we see that there is a spin-fluctuation peak in 
$\mbox{Im} \chi$ at a frequency $\omega_{max}(q)\simeq q v_F/S$. 
The peak height is of order $\mbox{Im} \chi_{max} \simeq S/\epsilon_F$. 
These estimates coincide with those 
found in conventional PM theory, and are in reasonable agreement 
with the available neutron data\cite{N}. 
In contrast, the correlation length of the 
ferromagnetic fluctuations is found in our approach to be of order 
$k_F\xi\simeq \sqrt{S_{eff}}\simeq \sqrt{R_W}$, and hence much shorter 
and less pressure dependent than in PM theory where 
$k_F\xi\simeq \sqrt{S}$. A direct study of this quantity would help to 
clarify the nature of spin fluctuations in liquid $^3\mbox{He}$. 
We also observe that in the whole domain where $q\ll k_F$, 
the $q^2/k_F^2$ terms in Eq.\ref{chidyn} is always negligible in 
front of $1-\tilde{I}/\epsilon_F^*$.
Hence, the spin fluctuation mode has always 
linear dispersion $\omega_{max}\propto q$, in contrast to PM theory 
where a regime $\omega_{max}\propto q^3$ is reached above 
a characteristic wavevector $\propto k_F/\sqrt{S}$ and below a low-energy 
scale $\propto \epsilon_F/S^{3/2}$. Here in contrast, a single energy scale 
exists $T_F^*\simeq\epsilon_F^*\simeq .66 R_W \epsilon_F/S$ 
($\simeq 700$mK at $p=0$, $\simeq 400$mK at $p=34$bars).
As a result, the 
susceptibility is {\it not} predicted to display a $1/T^{4/3}$ dependence 
at high temperature as in PM theory but rather follows a Curie 
law $\chi\propto 1/T$, in better agreement with experiment. The 
low-temperature behavior of the susceptibility however is, rather 
remarkably, predicted to be quite similar to the PM result even 
though Stoner mean-field theory applies to the low-energy quasiparticles. 
Indeed, the latter yields a low-temperature correction: 
$\epsilon_F^*\chi = S_{eff} [1- c S_{eff} (T/T_F^*)^2]$ so that: 
$\chi/\chi_0 = S [1- c'/R_W\ (ST/T_F)^2]$. 
This expression obeys the scaling $T\chi(T)=f(T/T_F^*)$ (since $R_W(p)$ 
weakly depends on pressure) and is similar to the PM result \cite{BMF}, 
which is in good agreement with experiments and differs from 
the result of naive Stoner theory.    
 
Finally, we briefly discuss the magnetic field dependence of the 
magnetization, which has been proposed as a way to discriminate between 
the ``almost ferromagnetic'' and ``almost localized'' 
approaches \cite{PN}. Recently, 
a remarkable experiment \cite{WWP} (at $p=26 \mbox{bars}$) 
has revealed that the metamagnetic behavior 
predicted within the GA \cite{DV} at 
$h_c^{GA}\simeq 25 \mbox{T}$ 
is not observed, up to (effective) fields of the order of $200 T$. In the 
present model, the magnetization as a function of field can be deduced 
from the corresponding result for the Hubbard model $m=f_H(h)$, by solving 
the equation:   
$m=f_H(h+Im)$. The function $f_H$ has been computed numerically in a 
previous work\cite{LGK}.
Using the parameters $(U,I)$ corresponding to $p=26 \mbox{bars}$ we find
that model (\ref{ham}) does present at half-filling a metamagnetic 
transition, though at a much higher field $h_c\simeq 80 \mbox{T}$ 
than the GA result. 
This indicates that 
the constraint of half-filling is too strict to provide a 
reasonable lattice-gas description of the magnetization experiments. 
We have checked that extending the model to allow for a small 
concentration of vacancies $\delta \sim 8 \% $
(in the spirit of Ref. \cite{SGRUV}) allows a reasonable description 
of the experimental magnetization curve $m(h)$.

In conclusion , we have proposed a physical picture of normal
$^3$He as a ``Mott-Stoner'' liquid , which seems in qualitative
agreement with several experimental apsects.

We are grateful to W.Krauth, P.\-Nozi\`eres,  
C.Lhuillier, P.E. Wolf and M.T. B\'eal-Monod for help and discussions.  
This article is dedicated to the memory of Sir Nevil Mott.

\end{multicols}

\begin{references}
\bibitem{EPS} Throughout this paper, the energy scale $\epsilon_F$
stands for the inverse of the non-interacting
susceptibility $\epsilon_F\equiv 1/\chi_0$.
It is related to the $^3$He
Fermi energy by: $E_F (^3\mbox{He}) = 3 \epsilon_F/2$.
\bibitem{DOS} Quantitative calculations are made here for a
semi-circular d.o.s:
$D(\epsilon)=\sqrt{1-(\pi\epsilon/4\epsilon_F)^2}/(2\epsilon_F)$
\end{references}
\end{document}